\documentclass{Interspeech2024}
\usepackage[nolist]{acronym}
\usepackage{cite}
\usepackage[acronym]{glossaries}
\usepackage[implicit=false]{hyperref}

\newacronym{DNN}{DNN}{deep neural network}
\newacronym{FLOPs}{FLOPs}{floating point operations per second}
\newacronym{ASR}{ASR}{automatic speech recognition}
\newacronym{LLM}{LLM}{large-language models}
\newacronym{TTS}{TTS}{text-to-speech}
\newacronym{SED}{SED}{sound event detection}
\newacronym{SSL}{SSL}{self-supervised learning}
\newacronym{MSE}{MSE}{mean square error}
\newacronym{BCE}{BCE}{binary cross entropy}
\newacronym{SOT}{SOT}{serialized output training}
\newacronym{CSR}{CSR}{conversational speech recognition}
\newacronym{SNR}{SNR}{signal-to-noise ratio}
\newacronym{RIR}{RIR}{room impulse response}
\newacronym{VAD}{VAD}{voice activity detection}
\newacronym{WER}{WER}{word error rate}
\newacronym{SotA}{SotA}{state-of-the-art}
\newacronym{SSE}{SSE}{speech separation and enhancement}
\newacronym{cpWER}{cpWER}{concatenated minimum permutation WER}
\newacronym{MIMO-WER}{MIMO-WER}{multi-input multi-output WER}
\newacronym{MVDR}{MVDR}{minimum variance distortionless response}

\interspeechfinaltrue
\title{Generating Data with Text-to-Speech and Large-Language Models for Conversational Speech Recognition}

\name[affiliation={1}]{Samuele}{Cornell*}
\name[affiliation={2}]{Jordan}{Darefsky*}
\name[affiliation={2}]{Zhiyao}{Duan}
\name[affiliation={1}]{Shinji}{Watanabe}

\address{
  $^1$Carnegie Mellon University, USA\\
  $^2$University of Rochester, USA}
\email{samuele.cornell@ieee.org, jdarefsk@u.rochester.edu}

\keywords{generative synthetic data, multi-talker speech recognition, text-to-speech,  conversational speech processing}

\begin{document}

\maketitle
\def\thefootnote{*}\footnotetext{These authors contributed equally to this work.}\def\thefootnote{\arabic{footnote}}

\begin{abstract}


Currently, a common approach in many speech processing tasks is to leverage large scale pre-trained models by fine-tuning them on in-domain data for a particular application. 
Yet obtaining even a small amount of such data can be problematic, especially for sensitive domains and conversational speech scenarios, due to both privacy issues and annotation costs. To address this, synthetic data generation using single speaker datasets has been employed. Yet, for multi-speaker cases, such an approach often requires extensive manual effort and is prone to domain mismatches. 
In this work, we propose a synthetic data generation pipeline for multi-speaker conversational ASR, leveraging a large language model (LLM) for content creation and a conversational multi-speaker text-to-speech (TTS) model for speech synthesis. We conduct evaluation by fine-tuning the Whisper ASR model for telephone and distant conversational speech settings, using both in-domain data and generated synthetic data. 
Our results show that the proposed method is able to significantly outperform classical multi-speaker generation approaches that use external, non-conversational speech datasets. 
\end{abstract}

\section{Introduction}

Current robust speech processing methods are considerably data hungry. For example, state-of-the-art \gls{ASR} systems require tens or even hundreds of thousands of hours of training data in order to achieve enough robustness in different domains~\cite{radford2023robust, peng2023reproducing, 
kanda2021large}. 
Such a vast amount of training data is leveraged either explicitly by training from scratch on a large amount of data or implicitly by fine-tuning/adapting a pre-trained ``foundation''  model that was originally trained, in a supervised or unsupervised manner~\cite{baevski2020wav2vec, hsu2021hubert, chen2022wavlm, radford2023robust}, on a large dataset. 

Nevertheless, for some domains, obtaining even a small portion of in-domain supervised data for fine-tuning can be problematic due to potential privacy concerns or prohibitive expense.

This is especially true for sensitive application scenarios, including medical, government, and law enforcement settings. Moreover, due to increasing regulatory attention, even scaling in-domain training data is potentially becoming more difficult. 

Aside from privacy issues,
applications that require recordings with multiple speakers are also inherently difficult, time-consuming and costly to annotate and thus obtain in scale. 
Prominent examples are meeting scenarios~\cite{watanabe2020chime, cornell2023chime} including doctor-patient recordings, speech captioning, speech analytics and so on.

Despite the difficulties associated with obtaining data for multi-speaker scenarios, there are speech processing approaches that require multi-speaker conversational data for training. These approaches have proven to be effective as demonstrated in recent speech processing challenges~\cite{watanabe2020chime,
cornell2023chime, ryant2021third}. Prominent examples are end-to-end neural diarization (EEND) and most target speaker voice activity detection (TS-VAD) approaches~\cite{fujita2019end, kinoshita2021integrating, landini2022simulated, medennikov2020target, 
tawara2023ntt}, as well as multi-speaker ASR~\cite{kanda2020serialized, kanda2021investigation, huang2023adapting, cornell2024one, liadapting}. Lack of annotated in-domain conversational data at scale is a significant issue for these techniques, which is only partly mitigated by leveraging foundation models~\cite{huang2023adapting, cornell2024one, liadapting}. 
Consequently, many of these approaches have to rely on synthetic data to increase dataset size. This is commonly achieved by artificially overlapping clips from existing datasets and adding noise and reverberation. 


While several toolkits have been proposed to ease the workload ~\cite{cordlandwehr2022mms_msg, park2023property}, creating synthetic datasets remains more art than science, as it often requires extensive hand-tuning, domain knowledge, heuristics, and significant trial and error. Crucially, this process is highly prone to the introduction of unwanted biases in the resulting dataset, leading to a performance drop due to domain mismatch~\cite{landini2022simulated}.

The aforementioned difficulties motivate the development of more automated, machine learning based approaches for synthetic data creation.
Several methods have in fact explored this direction, primarily focusing on improving \gls{ASR} performance by leveraging synthetic data created with \gls{TTS} models~\cite{rosenberg2019speech, chen2020improving, rossenbach2020generating, tjandra2020machine, fazel2021synthasr, zheng2021using, ueno2021data,zheng2021using, hu2022synt++, soleymanpour2022synthesizing, casanova2023asr} or leveraging \gls{ASR} and \gls{TTS} cycle-consistency during training~\cite{hori2019cycle, baskar2021eat} for semi-supervised training.
However, these approaches focus on single-speaker scenarios and thus cannot be directly applied to domains where multi-speaker conversational \gls{ASR} is required. 
In parallel, recent works~\cite{jung2024augsumm, wu2024improving} on speech summarization and audio captioning have shown how \gls{LLM}s can be leveraged effectively for synthetic audio data augmentation.

Building upon this previous research, in this work we explore using \gls{TTS} models along with \gls{LLM}s to generate multi-speaker conversational data. 
We focus on two-speaker \gls{ASR} on real-world telephone (Fisher~\cite{cieri2004fisher}) and distant speech recognition settings (Mixer 6 Speech~\cite{brandschain2010mixer}) by fine-tuning Whisper~\cite{radford2023robust}. 
The contributions of this work are the following: 1) We propose a synthetic data generation pipeline for conversational ASR using LLMs for content generation and a conversational multi-speaker TTS model for speech generation; 2) We perform a systematic investigation on the use of synthetic data for training multi-speaker ASR models with three different approaches:  using ``classical'' LibriSpeech based multi-speaker simulation, using a conventional \gls{SotA} \gls{TTS} model, and using a recently proposed conversational \gls{TTS} model~\cite{darefsky2024parakeet}.









\section{Method under study}\label{sec:method_under_study}

Our approach is summarized in Figure~\ref{fig:main_approach}. 
We explore the use of a pre-trained chat-optimized LLM for creating short conversation transcripts between two participants from scratch for when in-domain conversational transcriptions are not available or would be costly to obtain. 
Specifically, we use the recently released Llama 3 8B Instruct model and few-shot prompt it with 8 text-prompt examples randomly selected from a 1000-example subset of Spotify Podcasts dataset~\cite{clifton2020100} used to train Parakeet (the text data was transcribed using Whisper-D, described in ~\cite{darefsky2024parakeet}). That is, for each new example we want to generate, we randomly select a subset of eight text samples from our Parakeet subset to use as the few-shot prompt.
This procedure could also be used to augment existing in-domain text-only data. It could also be worth exploring fine-tuning on in-domain data instead of prompting.


These LLM obtained transcripts are then used to generate synthesized speech through a multi-speaker \gls{TTS} model. The resulting data, consisting of ground truth multi-speaker transcripts and the synthesized multi-speaker mixture can then be used for training or fine-tuning purposes, i.e. in Sec.~\ref{sec:experiments} for adapting Whisper to perform multi-speaker ASR. 






\subsection{Conversational TTS generation}\label{ssec:parakeet}

The effectiveness of this approach will heavily depend on the capability of the \gls{TTS} model used. While we expect \gls{LLM}s will be proficient in generating conversational transcripts as shown in previous work on summarization ~\cite{jung2024augsumm}, most \gls{TTS} models are not capable of synthesizing multi-speaker conversational data. Although one could naively generate each speaker’s utterances independently and then stitch them together, such an approach would fail to capture real conversational speech turn-taking dynamics and para-linguistic subtleties such as changes in intonation, etc., and would therefore potentially introduce a domain mismatch in the generated audio. 

\begin{figure}[h]
  \centering
\includegraphics[width=0.65\columnwidth]{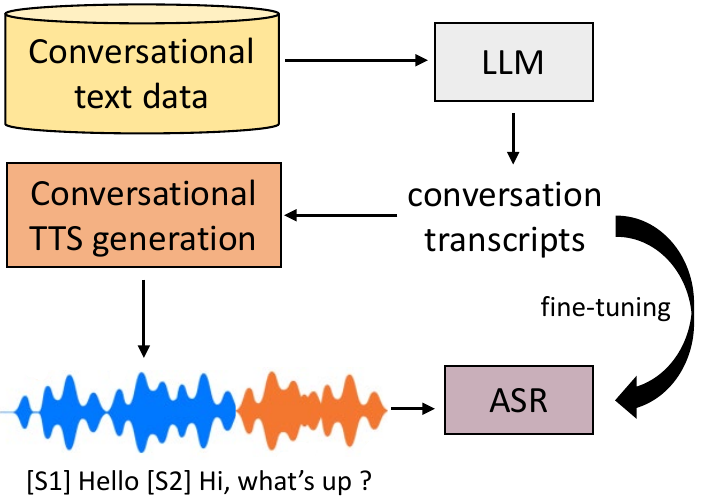}
  \caption{Block diagram of the proposed approach.}
  \label{fig:main_approach}
  
\end{figure}

Recently, in \cite{darefsky2024parakeet} a conversational \gls{TTS} model, Parakeet, has been proposed. Parakeet’s training dataset includes ~60,000 hours of Spotify Podcasts data, much of which is multi-speaker. It is therefore able to directly generate two-speaker short conversations of up to 30 seconds when given a text prompt in the style of the one in Figure~\ref{fig:main_approach}, i.e. with speaker-id related tags [S1] and [S2].
We use a diffusion version of Parakeet that, similar to~\cite{liu2024autoregressive} autoregressively generates blocks of continuous latents using latent diffusion on each block. The autoencoder is trained to map 44,100 Hz audio to 16-channel dimensional latents, with a time downsampling factor of 1024. Each diffusion block consists of 128 (time-wise) latent vectors, which correspond to approximately three seconds of audio.

LLM-generated transcripts and speech examples are available online\footnote{\href{https://popcornell.github.io/SynthConvASRDemo}{popcornell.github.io/SynthConvASRDemo}}.

\section{Experimental setup}

\subsection{Evaluation data}

In this work, we focus on two-speaker multi-speaker conversational ASR. 
This focus is due to the limitations of Parakeet, whose generations tend to lose correctness as the number of unique speakers in the text prompt increases. Furthermore, we also consider scenarios with relatively high \gls{SNR}; tackling more complex settings such as CH\-iME-6~\cite{watanabe2020chime} requires modeling of background noise and dynamic acoustic conditions (as the participants move, reverberation can change significantly). 
We thus perform our experiments using two conversational speech datasets with these characteristics: Fisher Corpus (both Part 1 and Part 2) and Mixer 6 Speech. 

\vspace{-0.5em}
\subsubsection{Fisher}\label{ssec:fisher_data}
Fisher consists of $11699$ telephone conversations between two English speakers sampled at $8$\,kHz. Each conversation is around $10$\,minutes long. 
We use the train, validation, and test split from~\cite{morrone2024end} ($11577$, $61$ and $61$ conversations of respectively $1960$\,h, $7$\,h and $7$\,h).   
The Fisher recordings originally separate each of the speakers into different channels;
however, since our focus is on general single-channel conversational speech processing, we mixdown the two channels to mono. We also resample the signal to $16$\,kHz as we use Whisper which was trained on $16$\,kHz data (see Sec.~\ref{ssec:asr_sys}).

\vspace{-0.5em}
\subsubsection{Mixer $6$ Speech}

As an additional scenario, we consider Mixer 6 Speech. Specifically we use the version re-annotated for the CHiME-7 challenge~\cite{cornell2023chime}. 
It consists of two-speaker interviews of approximately $15$ minutes (sampled at 16\,kHz) recorded by 14 different far-field recording devices. 
In this work we only use recordings from the tabletop microphone device (CH04). 
We use the splitting from~\cite{cornell2023chime}, where full long-form annotation is only available for the development ($59$ interviews, $15$\,h) and evaluation sets ($23$ interviews, $13$\,h). 
Here we further split the development set into an adaptation portion and a validation portion of respectively 2:30\,h and  
4\,h after discarding utterance groups longer than 30\,s as done in~\cite{liadapting}. 
This further split allows us to compare the use of synthetic data versus in-domain data for fine-tuning. 



\subsection{Baseline Methods}

\subsubsection{NeMo multi-speaker simulation tool}
We consider two baseline methods. The first method we consider is a ``classical'' synthetic speech generation method, where single speaker speech from one high quality speech dataset (e.g. LibriSpeech~\cite{panayotov2015librispeech}) is used to construct conversation-style synthetic recordings by artificially overlapping single speaker utterances and contaminating them by adding noise, artificial \gls{RIR} or other transforms (e.g. clipping, microphone transfer function etc.).  
We make use of the \gls{SotA} NeMo multi-speaker simulation tool~\cite{park2023property} (NeMo MSS in the following). 
We use LibriSpeech train-clean $360$ and $100$ portions and generate $100$\,h of short conversations between two speakers of up to $30$ seconds in length. 
For Mixer 6 Speech experiments, we additionally use the built-in \gls{RIR} simulation in order to generate simulated far-field speech. 

\subsubsection{xTTS-v2}
The second baseline method we consider is the approach outlined in Section~\ref{sec:method_under_study}, where a standard \gls{TTS} model is used to generate the training data. 
We explore this using the Coqui xTTS-v2 model~\cite{casanova2024xtts} (denoted simply as xTTS in Sec.~\ref{sec:experiments})
In detail, for each utterance group in the training dataset (either LLM-generated or taken from a text-only corpus) we sample two speaker ids from LibriSpeech train-clean $360$ and $100$ and then two corresponding LibriSpeech enrollment utterances to condition xTTS-v2 for the generated TTS id.  
We then generate each utterance in the utterance group independently via xTTS-v2 and truncate excessive leading and trailing silence regions using Silero VAD~\cite{SileroVAD}. The generated audio is then resampled to $16$\,kHz and mixed together by randomly adding start time offsets based on the order of the sentences in the utterance group transcript, ensuring that utterances from the same speaker do not overlap. 

\subsection{\gls{ASR} System}\label{ssec:asr_sys}

In our experiments, which focus on two-speaker conversational speech, we use the method proposed in ~\cite{liadapting} where Whisper~\cite{radford2023robust} is adapted to perform multi-speaker ASR through fine-tuning with a \gls{SOT}~\cite{kanda2020serialized} objective on utterance groups. 
This approach aligns with common practices in the field where a model pre-trained on a large amount of data (i.e. a foundation model) is fine-tuned/adapted for a particular domain or application of interest. 

Compared to~\cite{liadapting}, in our experiments we focus only on standard \gls{SOT} without considering timestamps and use only Whisper medium.  
We use low-rank adapters (LoRA)~\cite{hu2021lora} while the rest of the model is kept frozen.
During each fine-tuning experiment a linear warm-up schedule is employed for the first $N$ epoch, then the learning rate is linearly decayed over a maximum of $20$ epochs. The $L^2$ norm of the gradients is clipped to $5$. 
One LoRA adapter for each linear layer in the model (i.e. for each query, key, value and feed-forward network layer) is used. For each adapter we set the LORA rank to $64$, alpha to $128$, and dropout to $0.1$. 
In our preliminary experiments on the full Fisher training set, we found that this configuration yields the best results, even when compared to fine-tuning the entire model.
If validation loss does not improve for $2$ consecutive epochs the training is stopped.
We tune the batch size, number of warm-up epochs ($N$) and the value of the maximum learning rate for each set of experiments. Parakeet synthesized audio is resampled to $16$\,kHz in our experiments. In Fisher experiments, for all synthetic data, we use on-the-fly resampling to simulate telephone $3400$\,Hz band-limiting. 
\\ In Mixer 6 experiments, only for xTTS and Parakeet, we contaminate the data with reverberation using random RIRs obtained from~\cite{ko2017study}. This of course is less realistic than the \gls{RIR} simulation used in NeMo MSS as the \gls{RIR} is the same for both speakers.
We make our fine-tuning code publicly available\footnote{\href{https://github.com/popcornell/ASRLightningFT}{github.com/popcornell/ASRLightningFT}}. 

\subsection{Evaluation Setup}
For each dataset, we run our experiments using the same setup as in~\cite{liadapting}, where oracle \gls{VAD} is used and the dataset is divided into several utterance groups~\cite{kanda2021large, liadapting}. 
Continuing to follow~\cite{liadapting}, we then perform evaluation for each utterance group independently and accumulate \gls{WER} statistics over the whole dataset (insertions, deletions etc.). 
We choose this evaluation method because we only focus on multi-speaker ASR, and an evaluation which considers the whole conversation (e.g. as in CHiME-6/7) would require a diarization component, which would add significant complexity. 

We thus consider \gls{cpWER}~\cite{watanabe2020chime}. This is the same as \gls{WER} in~\cite{liadapting}, with the best permutation evaluated for each utterance group independently.
We also consider \gls{MIMO-WER}, which is more tolerant than \gls{cpWER} to speaker assignment errors. We use the Meeteval toolkit~\cite{vonmeeteval} to compute both scores.
Whisper text normalization is used both during training and scoring. 






\section{Experiments}\label{sec:experiments}

\subsection{Fisher}

In Table~\ref{tab:fisher} we report results obtained on the Fisher test set as defined in Sec.~\ref{ssec:fisher_data} with different data used for fine-tuning. 
As a baseline, in the first row, we report the results with no adaptation. In the second panel, we report results on in-domain Fisher training data adaptation. We observe only a modest difference between using the full training set or a 80\,h data subset, which is likely because we are leveraging a strong pre-trained model.
In the third and fourth panels, we report results obtained with synthetic data approaches. 
In particular, for the two \gls{TTS} approaches (xTTS and Parakeet), we consider two opposite situations: a best-case/oracle scenario where we use in-domain conversation transcriptions and another one where we suppose we have none and thus we use as input Llama-3 random generated utterance groups transcripts (LLM$_{rnd}$) as described in Sec.~\ref{sec:method_under_study}. 

We observe that xTTS-based generation outperforms NeMo MSS when Fisher only transcriptions (Fisher) are used. When LLM generated transcriptions are used (LLM$_{rnd}$), xTTS performance is on par/slightly worse than NeMo MSS.
In contrast, when using Parakeet, the difference between using LLM generated transcripts versus the Fisher training set transcriptions is modest, and interestingly, the generated transcripts yield the best performance. 
In general, while the performance gain compared to the baseline synthetic data approaches (xTTS and NeMo MSS) is significant, there remains a substantial gap compared to using in-domain data (Fisher). It appears that this gap cannot be bridged solely by scaling the amount of synthetic data.

\begin{figure}[t]
  \centering
\includegraphics[width=0.97\columnwidth]{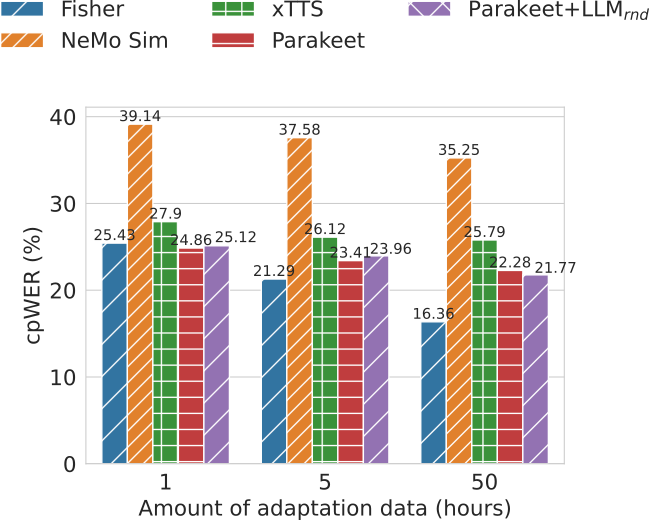}
  \caption{Multi-speaker ASR results on Fisher test set for different adaptation data sources and quantity.}
  \label{fig:fisher_barplot}
\end{figure}

In Figure~\ref{fig:fisher_barplot} we report cpWER on Fisher for different amounts of adaptation data, both from Fisher training set and from synthetic approaches. 
For modest amounts of data (less than 5\,h) the proposed approach is competitive to using in-domain data; however, as the amount of adaptation data is scaled, performance saturates quickly: The improvement between 50\,h and 5\,h is marginal when compared to the one afforded by using in-domain data. 
This trend is also observed for the other synthetic data approaches and suggests that there is some inherent mismatch in all of the synthetic data approaches tested that prevents effective scaling. 
At least for Parakeet, results suggest that this mismatch seems to be more related to the signal/acoustic content rather than the transcription semantic content as the gap between using Fisher transcriptions and LLM-generated transcription is modest. 


\begin{table}[]
\centering
\caption{Multi-speaker ASR results on Fisher test set with different adaptation data.}
\vspace{-0.2cm}
\label{tab:fisher}
\setlength{\tabcolsep}{2.1pt}
\begin{tabular}{lccc}
\toprule
Adaptation Data & amount & cpWER  & MIMO-WER \\
      & (hours) & (\%) & (\%) \\
 \midrule
 -  & 0 & 44.94 &  26.15 \\

\midrule
Fisher & $1960$ & 13.76  &  13.58 \\ 
 Fisher & $80$ & 15.43  &  14.94 \\
\midrule
 NeMo MSS & $80$ & 34.37 &  26.51 \\ 
 xTTS (Fisher) & $80$ & 24.88 & 24.07 \\
 xTTS (LLM$_{rnd}$) & $80$ & 34.65 & 28.31 \\
\midrule
 Parakeet (Fisher)  & $80$ & 21.44 & 21.00 \\
 Parakeet (LLM$_{rnd}$)  & $80$ & 20.41 & 19.48 \\
 Parakeet (LLM$_{rnd}$)  & $160$ & 19.93 & 19.45 \\

\bottomrule
\end{tabular}
\end{table}

\subsection{Mixer 6 Speech}

In Table~\ref{tab:mixer6}, we show results obtained on Mixer 6. 
The trends observed are consistent with the Fisher experiments, despite the rather naive artificial reverberation strategy used for xTTS and Parakeet experiments. 
This confirms that the proposed approach can also be effective for far-field multi-speaker synthetic data, at least when compared to the classical approach (NeMo MSS results) and when available in-domain data is very scarce (here 2:30\,h).
Parakeet (LLM$_{rnd}$, 80\,h) also compares favorably with the third and fourth rows, where we report the results of using the Fisher full 1960\,h training set and a 80\,h subset respectively for adaptation. For these Fisher experiments, to reduce the mismatch due to the telephone lower sampling frequency, we apply telephone band-limiting to Mixer 6 in the inference phase. We also contaminate the Fisher 6 training data with reverberation as done for Parakeet and xTTS as described in Sec.~\ref{ssec:asr_sys}.

\subsection{Further discussion \& remarks}

Considering both Fisher and Mixer 6 experiments, the fact that Parakeet+LLM$_{rnd}$ improves considerably over NeMo MSS while xTTS fails suggests that turn-taking and para-linguistics may play a considerable role for multi-talker ASR. 

Finally, for both Mixer 6 Speech and Fisher scenarios, we tried using $50\,h$ of synthetic LLM$_{rnd}$ data to augment a portion of in-domain data ($5\,h$ and $50\,h$) by mixing the two or by training on synthetic data and then fine-tuning on in-domain data. 
However, in most instances, this approach does not result in any improvement over using solely the in-domain data; in the xTTS and NeMo MSS cases we even observe performance degradation.
For example, by combining  $50\,h$ of Parakeet (LLM$_{rnd}$) and $50\,h$ of original Fisher training data the model achieved a cpWER of $15.74$\% which is only marginally better than the $16.36$\% obtained with only $50\,h$ of Fisher (Figure~\ref{fig:fisher_barplot}). 
Interestingly, negligible or no improvement was also observed when the in-domain data was more modest (5\,h). 
This may be due to the fact that we are leveraging a strong pre-trained model, and thus the quality of adaptation data rather than quantity matters most. 
Future work should explore adaptation of the \gls{TTS} model to generate synthetic audio that better matches distribution of in-domain data.


\begin{table}[]
\centering
\caption{Multi-speaker ASR results on Mixer 6 Speech eval set with different adaptation data.}
\vspace{-0.2cm}
\label{tab:mixer6}
\setlength{\tabcolsep}{2.1pt}
\begin{tabular}{lccc}
\toprule
Adaptation Data & amount & cpWER & MIMO-WER  \\
  & (hours) &  (\%)  &   (\%)\\
\midrule
 -  & 0 & 43.67  &  32.16  \\
\midrule
Mixer6 & 2.30 & 20.36 & 19.77  \\ 
Fisher  & 1960 & 20.83 & 20.33 \\
Fisher  & 80 & 22.12 & 21.36 \\
NeMo MSS  & 80 & 36.71 & 28.21 \\
xTTS (Mixer6) & 2.30 & 25.99  &  24.47  \\
xTTS (LLM$_{rnd}$) & 80 &  35.65 & 30.18  \\
\midrule
Parakeet (Mixer6) & 2.30 & 23.52 & 22.82 \\

Parakeet (LLM$_{rnd}$) & 2.30 & 23.70 & 22.12 \\
Parakeet (LLM$_{rnd}$) & 80 & 21.25 & 20.17  \\

\bottomrule
\end{tabular}
\vspace{-3mm}
\end{table}


\section{Conclusions}
In this work, we study the use of synthetically generated data for multi-speaker \gls{ASR}, focusing on the two-speaker case. 
We explore different strategies of generating synthetic data, comparing artificially overlapped data and \gls{SotA} conventional \gls{TTS} models with a novel conversational \gls{TTS} model, Parakeet, capable of natively generating multi-speaker utterances. 
Our results show that our approach using Parakeet significantly outperforms previous \gls{SotA} multi-speaker simulation techniques. 
Furthermore, when in-domain data is limited to only a few hours, our approach achieves performance reasonably close to that of using in-domain data; however, when more in-domain data is available, our approach lags behind using real data. For Mixer 6, our approach also obtains results comparable to using external real-world multi-speaker data (Fisher). 
Overall, our experiments suggest that the LLM generated transcripts are reliable but that there is currently a performance gap compared to using in-domain audio data (when enough in-domain data exists).

Limitations of our work include that we only consider two-speaker conversational speech, short 30-second conversations, and relatively high SNR scenarios. These constraints were primarily imposed by the current limitations of the Parakeet TTS model, and thus improvement of TTS capabilities is crucial to increasing synthetic data viability. For example, to tackle more complex noisy/reverberant scenarios, the TTS model needs to incorporate acoustic scenario modeling, e.g. via acoustic style transfer techniques or even few-shot adaptation on some in-domain data (e.g. via~\cite{zhang2023adding}). Another possible limitation is that Parakeet itself is trained on text-audio pairs where the text is ``synthetic'', i.e. Whisper-D~\cite{darefsky2024parakeet} is used to generate multi-speaker transcriptions for Spotify podcast audio which is then used for Parakeet training. Since Whisper-D is fine-tuned from Whisper using a small number of annotated multi-speaker examples (and Whisper itself is likely trained on a sizeable quantity of multi-speaker data), there is an indirect but somewhat circular dependency on the existence of ground-truth annotations. Also, Parakeet's weakness in generating consistent 3/4-speaker conversational data could in part be due to limitations of Whisper-D. Future work could potentially explore the joint bootstrapping of audio-to-text and text-to-audio models.



\section{Acknowledgments}
S. Cornell was supported by IC Postdoctoral
Research Fellowship Program at CMU via ORISE through an agreement between U.S. DoE and ODNI. 
We'd like to thank Google's TPU Research Cloud (TRC), which provided compute for generating synthetic Parakeet samples and Llama synthetic text utterances. Our work would not have been possible without their support.

\bibliographystyle{IEEEtran}
\footnotesize
\bibliography{mybib}

\end{document}